\documentclass[10pt, conf, twocolumn]{IEEEtran}

\usepackage{color}
\usepackage{times}
\usepackage{epsfig}
\usepackage{graphicx}
\usepackage{epstopdf}
\usepackage{algorithm}
\usepackage{algpseudocode}
\usepackage{amsmath}
\usepackage{amssymb}
\usepackage{bbm}
\usepackage{amsxtra}
\usepackage{multirow}
\usepackage{mathtools}
\usepackage{caption}

\usepackage{url}
\usepackage{float}
\usepackage[font=small, singlelinecheck=false]{caption}
\usepackage{caption}
\usepackage[caption = false]{subfig}
\usepackage{cite}
\usepackage{amsthm}
\newtheorem{theorem}{Theorem}
\newtheorem{lemma}{Lemma}

\newtheorem{corollary}{Corollary}

\newtheorem*{proof*}{Proof}
\usepackage{adjustbox}
\usepackage{tikz}
\usepackage{hyperref}
\usepackage{cleveref}

\begin{document}

\title{Real-Time Transmission Mechanism Design for Wireless IoT Sensors with Energy Harvesting under Power Saving Mode}

\author{ \IEEEauthorblockN{\large Jin Shang, Muhammad Junaid Farooq,~\textit{Student Member, IEEE}, and Quanyan Zhu,~\textit{Member, IEEE}} \vspace{-0.0in}
\thanks{\vspace{-0.1in}\hrule \vspace{0.1cm}
Jin Shang is with the Department of Mathematics, New York University Abu Dhabi, UAE. Muhammad Junaid Farooq and Quanyan Zhu are with the Department of Electrical \& Computer Engineering, Tandon School of Engineering, New York University, Brooklyn, NY 11201, USA, E-mails: \{js8544, mjf514, qz494\}@nyu.edu.
}

}

\maketitle

\begin{abstract}
The Internet of things (IoT) comprises of wireless sensors and actuators connected via access points to the Internet. Often, the sensing devices are remotely deployed with limited battery power and are equipped with energy harvesting equipment. These devices transmit real-time data to the base station (BS), which is used in applications such as anomaly detection. Under sufficient power availability, wireless transmissions from sensors can be scheduled at regular time intervals to maintain real-time data acquisition. However, once the battery is significantly depleted, the devices enter into power saving mode and need to be more selective in transmitting information to the BS. Transmitting a particular piece of sensed data consumes power while discarding it may result in loss of utility at the BS. The goal is to design an optimal dynamic policy which enables the device to decide whether to transmit or to discard a piece of sensing data particularly under the power saving mode. This will enable the sensor to prolong its operation while causing minimum loss of utility to the application. We develop an analytical framework to capture the utility of the IoT sensor transmissions and leverage dynamic programming based approach to derive an optimal real-time transmission policy that is based on the statistics of information arrival, the likelihood of harvested energy, and designed lifetime of the sensors. Numerical results show that if the statistics of future data valuation are accurately predicted, there is a significant increase in utility obtained at the BS as well as the battery lifetime.
\end{abstract}

\IEEEpeerreviewmaketitle


\section{Introduction}
\begin{figure}[t]
    \centering
    \includegraphics[width=\columnwidth]{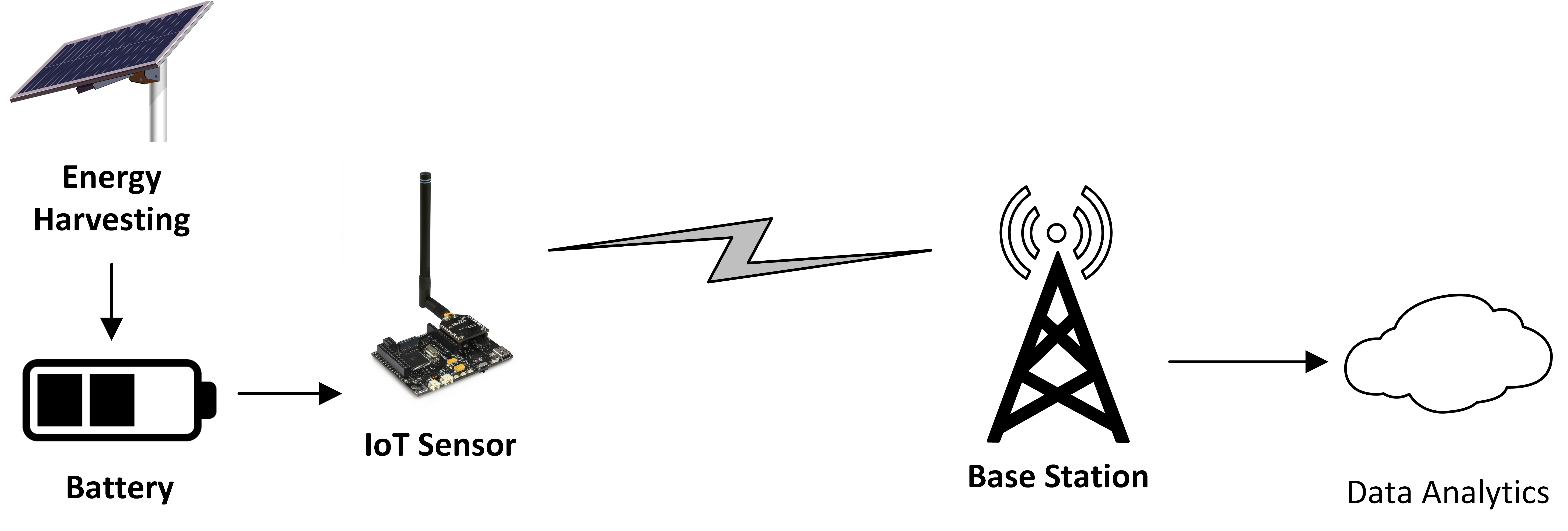}
    \caption{Illustration of an uplink IoT system. A battery powered sensor equipped with energy harvesting device transmits sensed data to the base station, which is subsequently used for data analytics.}
    \label{fig:system_model}
\end{figure}

The Internet of things (IoT) comprises of wireless sensors and actuators connected via access points to the Internet.
The proliferation of IoT devices 
is enabling tremendous applications and use cases in sectors such as transportation, healthcare, smart cities, and infrastructure systems, etc. ~\cite{iot_fuqaha}. Many IoT applications in these scenarios such as waste water monitoring, smart garbage collection, smart street lighting systems, air quality monitoring, asset tracking, etc., require continuous low-rate streaming data reported to a centralized location from battery powered devices over a long period of time~\cite{lora_survey2}. Often, the devices are deployed at locations where mains power delivery is difficult or not possible. Moreover, it is difficult and costly to frequently replace batteries of sensors at remotely deployed devices. Therefore, the trend is towards using devices with energy harvesting capabilities coupled with employing power saving techniques to extend the operational lifetime of the devices to the order of several years~\cite{green_iot,iot_energy_efficient}. An illustration of the IoT sensing ecosystem under consideration is provided in Fig.~\ref{fig:system_model}.

Since sensors are usually required to continuously transmit real-time sensed data to a base station (BS), which is then converted into actionable insights, there is a persistent strain on the battery that reduces the operational lifetime of the sensors. One of the major sources of battery depletion is the uplink wireless transmissions by the IoT sensor. Therefore, the key to reducing the energy consumption and subsequently extending battery lifetime in IoT devices is to effectively control the wireless transmissions in order to achieve sensing goals while using the least amount of battery power\footnote{The frequency of wireless transmissions is one of the factors that can be modified to significantly alter the battery lifetime.}. In other words, more energy-efficient medium access control (MAC) protocols need to be developed to achieve these goals. To this end, some energy efficient MAC protocols have emerged, referred to as low power wide area networks (LPWANs)~\cite{lpwan}. One such protocol developed for IoT devices is known as the long range wide area network (LoRaWAN) protocol~\cite{lorawan}, which is optimized for battery-powered end-point devices. These MAC protocols allow for the remotely deployed devices in LoRa networks to operate for several years without the need for replacing the battery~\cite{lora_networks}.

Three different classes of LoRa~\cite{lora_survey2} devices have been developed namely Class A, B, and C, out of which Class A and B devices are battery powered while Class C are mains powered~\cite{lorawan_comparison}. These protocols reduce energy consumption by restricting the wireless transmission and reception in fixed time windows since RF transmissions cost the most in terms of energy in all classes of the LoRa protocol~\cite{lora_mac_hassaneen}. Class A devices only have two short receive windows after transmitting a packet. After the receive windows, Class A devices goes to sleep for conserving energy. Class B devices, on the other hand, have extra receive windows at scheduled intervals~\cite{lorawan_comparison}. While these protocols have been shown to be successful in many practical scenarios, there are several shortcomings of such periodic schemes. Firstly, periodic transmissions may result in missed or delayed reception of critical real-time data at the BS, which may have been generated outside of the scheduled transmission period. Secondly, sending data at regular intervals may be unnecessary as all sensor measurements might not be valuable enough to be transmitted. Moreover, in real-time applications, the data transmissions cannot be delayed until the start of the transmission period. Hence, there is a need to incorporate the importance of sensed data into the transmission decisions of MAC protocols.


While research is being carried out on improving the physical layer efficiency of these MAC protocols, we propose a cross-layer design for adding more intelligence on top of the protocol. This is possible because more and more powerful computational capabilities are being embedded onto IoT devices. In this paper, we consider real-time mission-critical IoT applications, which cannot tolerate delay in the data transmission. In other words, sensed data needs to be sent to the BS immediately after generation and a significant delay would make the transmission futile. The problem under consideration is to design a policy for real-time transmission decision based on the potential utility of transmitting the data. Particularly, the cognitive framework is useful once the device enters into power saving mode to prolong its period of operation without requiring a battery replacement. We develop a framework that uses knowledge of the amount of remaining battery, the usefulness of sensed data, and the channel state information to effectively decide on transmitting or discarding generated data. The goal is to improve the expected utility obtained from transmitting data by being more selective and adaptive in the transmission decisions. We leverage theories from dynamic programming and stochastic processes to construct an optimal dynamic transmission policy framework for use under a power saving mode in wireless IoT sensors.







\subsection{Related Work}
Power saving has been one of the key focus areas for researchers working on wireless sensors and IoT endpoint devices~\cite{power_saving_iot}. Recently, the Defense Advance Research Projects Agency (DARPA) has started a Near Zero Power RF and Sensor Operations (N-ZERO) program that is aimed at developing sensors for military purposes, which can stay dormant with negligible power consumption until awakened by an external trigger~\cite{n_zero}. Apart from reducing the power consumption of device circuitry, it requires the development of more cognitive MAC protocols to support the low power operation.
In order to reduce energy consumption in conventional cellular networks, sleep modes have been designed for existing Long Term Evolution (LTE) systems~\cite{sleep_mode_design}. Specifically, for continuously transmitting devices, a new mechanism referred to as discontinuous reception/transmission (DRX/DTX)~\cite{drx} has been incorporated into the 3GPP LTE-A standard, which allows devices to go to sleep by turning off their radio interfaces in different patterns. A significant amount of work is available in the literature that optimizes these sleep-wake patterns by considering the quality of service constraints~\cite{sleep_schedule_optimization}. Generally, the alternative sleep/wake cycles are proposed for delay tolerant applications where the data is buffered to be transmitted when the sensor is in the active stage~\cite{power_saving_scheduling}. Research in this direction focuses on designing energy efficient sleep schedules for IoT devices to achieve a target QoS and consider metrics such as bit-rate, packet delay, and packet loss rate. While these metrics are important for continuous data transmissions, in most IoT networks, the data is transmitted only at the occurrence of an event and hence such metrics may not be meaningful from a practical standpoint. Other works propose online and offline transmission policies for traditional communication data without considering the importance of transmitting it. They aim to maximize the long term average throughput by adjusting the transmit power of the transmitter in real-time~\cite{ulukus1}. The eventual goal is to transfer maximum amount of data from the source to the destination with minimum power consumption~\cite{ulukus2,yener_procrastinating_policy}. However, in our work, the objective is to transmit the most valuable information to the BS.


Existing approaches towards achieving energy efficiency in data transmission is limited to effective scheduling of transmissions and aligning them with the harvested energy. Fewer efforts have been made towards real-time transmission mechanism considering the energy efficiency. While offline transmissions can be scheduled according to the harvested energy, the real-time data needs to be transmitted immediately. Otherwise, it loses the utility. In this paper, we propose a data centric framework for real-time transmission decisions in the uplink of IoT sensors. The allocation of stochastic sequentially arriving tasks to resources has been well studied in management science and operations research literature~\cite{dlr}. Dynamic allocation of resources to stochastic and sequentially arriving requests has been studied in the cloud computing context in~\cite{junaid_acc}. However, it is based on a continuous time model as the requests arrive at random times. We leverage the developed models and methodologies to make online transmission decisions and expand it to encompass energy availability and wireless channel uncertainties.

\subsection{Contributions}
In this paper, we consider a dynamic transmission decision problem where sensing data is generated continuously with stochastic valuation and is transmitted over an uncertain wireless channel with limited randomly harvested energy. A summary of the main contributions is provided below:
\begin{enumerate}
\item We propose a optimal decision framework for transmission, based on the statistics of the importance of sensed data and instantaneous channel conditions, that maximizes the total expected utility in the presence of limited energy storage.
\item  We provide a dynamic decision support algorithm that can be computed offline and is implementable in real-time scenarios.
\item  We provide numerical evaluation of the performance of the optimal transmission mechanism and provide comparison with existing approaches.
\end{enumerate}

The rest of the paper is organized as follow: Section~\ref{Sec:Syst_Model} provides a description of the system model and its assumptions. Section~\ref{Sec:Method} provides a systematic methodology for optimally solving the transmission decision problem. Section~\ref{Sec:Results} contains results from the numerical evaluation of the framework for some special cases and the comparisons with existing approaches. Finally, Section~\ref{Sec:Conclusion} concludes the paper with a discussion on the future research directions.

\section{System Model} \label{Sec:Syst_Model}

\begin{table}[]
    \centering
    \caption{List of notations with description.}
    \setlength\tabcolsep{3pt}
    \label{parameters}
    \begin{tabular}{|l|l|}
        \hline
        Symbol & Description                                                             \\ \hline
        $N\in \mathbb{Z^+}$ & Number of available battery units               \\ \hline
        $n\in \mathbb{Z^+}$         &   Number of target measurements remaining     \\ \hline
        $x_i \in \mathbb{R}$&   Valuation of the measurement in the $i^{\text{th}}$ time slot                 \\ \hline
        $f_X(x)$ & Probability density function of sensed data valuation\\ \hline
        $F_X(x)$ & Probability distribution function of sensed data valuation\\ \hline
        $\mathcal{P}_s^i \in [0,1]$&   Probability of transmission success from sensor to BS  \\ \hline
        $U(.) \in \mathbb{R}$&   Utility obtained by successful transmission of sensed data                  \\ \hline
        $\pi \in [0,1]$ & Probability of harvesting a unit of energy in each time slot\\ \hline
        $h \in \mathbb{R}$ & Fast fading wireless channel gain\\ \hline
        $\rho_{\text{th}} \in \mathbb{R}$ & Channel quality threshold\\ \hline
        $\alpha_1$ & Probability of transmission success under high channel gain \\ \hline
        $\alpha_0$ & Probability of transmission success under high channel gain \\ \hline
        $g(\pi_{i}|\pi_{i-1})$ & Distribution of  $\pi_i$ conditional on the previous value \\ \hline
        $V_{N,n}(\pi,x)$ & The expected maximum total utility function \\ \hline
    \end{tabular}
\vspace{-0.0in}
\end{table}

In this section, we provide a description of the network model. For the convenience of readers, the notations used throughout this paper are summarized in Table~\ref{parameters} along with a brief description.

\vspace{-0.0in}
\subsection{System Model}
We consider an uplink IoT sensing system comprising of a sensor and a BS separated by a fixed distance. The sensor transmits sensed data to the BS in realtime, i.e., does not store it locally to be transmitted at a later period in time. Moreover, the sensor is equipped with energy harvesting capability to harness energy from renewable sources. We consider a time slotted system, in which sensing data is generated at intervals of duration $\tau$ s. Sensors are equipped with limited battery, whose capacity can be characterized with reference to the energy consumption of a single uplink transmission\footnote{We assume that the energy consumption of circuit processing and idle operation of the sensor is negligible as compared to the power consumption of uplink transmission.}. A target operation time, denoted by $T$ is set for the sensor, which implies that $n = \lfloor \frac{T}{\tau} \rfloor \in \mathbb{Z}^+$ is the target number of measurements that will be made by the sensor. We consider that the system is in power saving mode, i.e., there are $N$ units of battery energy remaining and $n \geq N$ measurements to be made in order to achieve the target lifetime of the sensor. Under normal operation mode, i.e., when sufficient amount of battery power is available, the sensor may apply standard or more customized transmission protocols optimized for other performance metrics such as false alarm rate, mis-detection probability, etc. However, once the power saving mode is active, then a more strategic mechanism for utility maximization is considered to be in place. Every uplink transmission costs the same amount of battery power. Upon successful receipt of the data, a utility is obtained by the BS.



At each time instant, sensed data is generated by the sensor and based on the measurement, its \emph{valuation} or significance for the application is determined. This valuation is denoted by $X\in \mathbb{R}^+$ and is modeled as a random variable with probability density function (pdf) $f_X(x)$ and cumulative distribution function (cdf) $F_X(x)$. The valuation of the data generated at the $i^{\text{th}}$ time slot is denoted by $x_i, i  = 1, \ldots, n,$ and is considered to be independent and identically distributed across the time slots.

\subsection{Channel Model}
The transmissions from the sensor to the BS experience wireless channel effects including signal attenuation and fading. We assume that the wireless channel experiences Rayleigh fading, i.e., the random channel gain, denoted by $\mathbf{h}$, follows an i.i.d. exponential distribution with mean $\mu^{-1}$ and $h_i$ denotes the channel gain realization at the $i^{\text{th}}$ time slot. We assume that complete channel state information (CSI) is available to the IoT device\footnote{This assumption is reasonable for practical systems since the device may be receiving discovery beacons from the BS. Moreover, there might be dedicated control channels available to convey such information.}. It implies that both instantaneous (or short term) CSI and statistical (or long-term) CSI is known to the sensor. The instantaneous CSI is used to make real-time decisions on transmissions while the expectation about future channel conditions comes from the statistical CSI.



At any particular time instant, the channel between the sensor and BS can be in either good or bad state based on the realization of the channel gain. Considering averaged effect of interference to the signal and the randomness in sensor-BS separation, the channel gain effectively plays the deciding role in whether a particular transmission will be successfully received or not. Therefore, we use a binary abstraction model that considers two states of the channel using a threshold on the channel gain denoted by $\rho_{\text{th}}$. The average probability of successful transmission during good and bad channel state is denoted by $\alpha_1$ and $\alpha_0$ respectively. These parameters incorporate the impact of the distance dependent signal decay as well as the interference and other channel effects.
Therefore, the probability of successful transmission\footnote{The probability of success is a compound metric that incorporates the effect of the path loss, channel gain, as well as the interference from other transmitting devices.} conditioned on the knowledge of the channel gain, denoted by $\mathcal{P}_s$ can be expressed as follows:
\begin{align}
\mathcal{P}_s =
    \left\{
	\begin{array}{ll}
		\alpha_1  & \mbox{if } h \geq \rho_{\text{th}}, \\
		\alpha_0 & \mbox{if } h < \rho_{\text{th}}.
	\end{array}
\right.
\end{align}
Note that the the probability of success is a random variable that depends on the state of the channel. The expected probability of successful transmission can be consequently expressed as follows:
\begin{align}
    \mathbb{E}[\mathcal{P}_s] &= \alpha_1 (1 - F_{h}(\rho_{\text{th}})) + \alpha_0 F_{h}(\rho_{\text{th}}), \\
    &= \alpha_0 + e^{-\mu \rho_{\text{th}}}(\alpha_1 - \alpha_0), \label{expected_P_s}
\end{align}
where~\eqref{expected_P_s} follows from the Rayleigh fading assumption on the channel.

\subsection{Energy Harvesting}
We consider the case of energy harvesting enabled sensors that harness available renewable energy to be used for transmission. At each time slot, a unit\footnote{The slot interval $\tau$ can be adjusted to ensure that a unit of energy can be generated that is sufficient to support a single transmission.} of renewable energy may be harvested with a probability $\pi \in [0,1]$ independently with other time slots. With a slight abuse of notation, we denote $N_i$ as the number of battery units available at the $i^{\text{th}}$ time slot. We assume a unit transition Markovian energy harvesting model, i.e., at each time slot, there is a probability that a unit of energy will be harvested. Therefore, the number of battery units available at the $i^{\text{th}}$ time slot evolve as follows:
\begin{align}\label{battery_dynamics}
    N_{i+1} =
    \left\{
	\begin{array}{ll}
		N_i + 1,  & \mbox{w.p.  } \pi, \\
		N_i, & \mbox{w.p.  } 1 - \pi.
	\end{array},
	i = 1, \ldots, n-1,
\right.
\end{align}
Note that once the battery is completely depleted, the sensor is assumed to shut down and any further harvested energy does not increase the operational lifetime\footnote{We assume that the sensor requires configuration with the BS initially and harvested energy may not be able to make the sensor operational once it has completely shut down.}.




\subsection{Realt-Time Transmission Decision Problem}


Upon successful reception of sensed data by the BS, a utility is obtained, which is defined $U(x_i): \mathbb{R} \rightarrow \mathbb{R}$. It is assumed to be invertible and monotonically increasing in the valuation $x_i$, i.e., a measurement with high valuation results in high utility at the BS.
The goal of the sensor is to strategically pick $N$ out of $n$ measurements and transmit them to the BS in order to maximize the total utility obtained using the battery energy remaining. The challenge in the selection lies in the fact that the decision has to be made in real-time without complete information about the valuation of future data arrivals. This can be mathematically expressed as follows:
\begin{align}
 V(N,n) = \ &\underset{\mathbf{y} \in \mathcal{Y}}{\text{maximize}}
& & \mathbb{E} \Big[ \sum_{k=1}^{n} U(x_{k}) \mathcal{P}_{s}^k y_k  \Big] \label{optim_obj} \\
& \text{subject to}
& & \text{Battery level dynamics in}~\eqref{battery_dynamics}. \label{optim_const} \\
& & & y_k \in \{0,1\}, \forall k = 1, \ldots, n. \label{optim_const2} \\
& & & \sum_{k=1}^n y_k = N. \label{optim_const3}
\end{align}
where $\mathcal{P}_s^k$ denotes the probability of successful transmission in the $k^{\text{th}}$ time slot. Hence, the vector $\mathbf{y}$ denotes the transmission policy, whereby $N$ out of $n$ measurements are selected to be transmitted. Note that the battery level at each time slot $N_i$ evolves stochastically according to the dynamics expressed in~\eqref{battery_dynamics}.
$\mathcal{Y}$ denotes the set of all possible permutations of the vector $[\mathbf{1}^{N}, \mathbf{0}^{n-N}]^T$.
In essence, the problem is to select the best data to transmit to the BS under when the data is arriving sequentially while considering the instantaneous channel quality.



\vspace{-0.0in}
\section{Methodology} \label{Sec:Method}

In this section, we provide the solution to sensor's optimization problem presented in~\eqref{optim_obj} to~\eqref{optim_const3}. In essence, the optimization entails to a stochastic dynamic programming problem with stage dynamics. Therefore, we can break the original problem into identical subproblems as described in the sequel.
At each time instant, the sensor is faced with a decision problem regarding whether it should transmit or discard the measurement. This decision is based on several factors such as the instantaneous CSI, the number of battery units available, the number of remaining measurements, the statistical information about the valuation of sensed data, and the statistical information of harvested energy.

\begin{figure*}[t!]
	\centering
	\vspace{0.1cm}
	\includegraphics[width=5in]{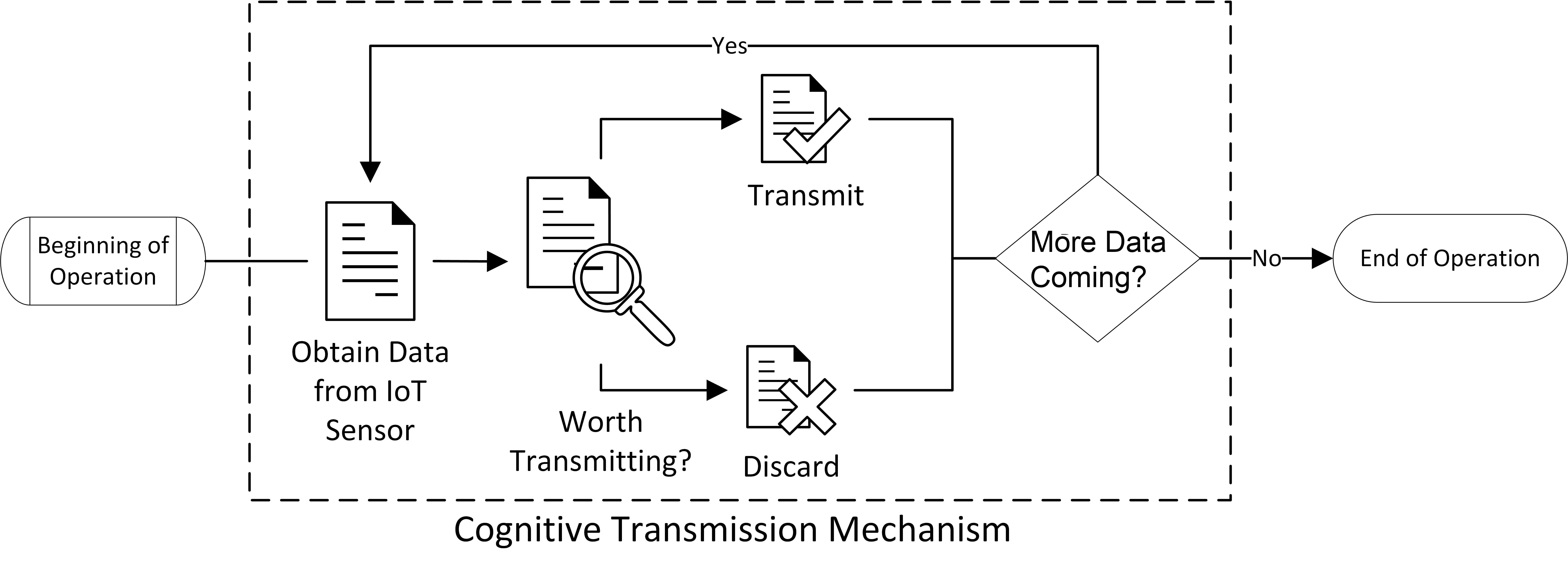}
	\caption{Cognitive process flow diagram of the realtime transmission mechanism. In every cycle, the sensed data is evaluated for its utility before being transmitted to the BS.\vspace{-0.1in}}
	\label{fig:flow}
\end{figure*}

\subsection{Optimization}

We denote the optimal value obtained by solving the optimization problem in~\eqref{optim_obj} to~\eqref{optim_const3} if  $n$ battery units are available and $N$ measurements are remaining, as $V (N,n)$.  Let $a_N^n \in  \mathbb{R}^+$ represent the decision threshold on the average utility of transmitting sensed data, i.e., $U(x_k) \mathcal{P}_s^k$ if $n$ battery units are available and $N$ measurements are remaining. Then, the value function can be expressed recursively using the following lemma:
\begin{lemma}\label{lemma_value_func}
	If $N$ battery units are remaining and $n$ measurements are targetted, then the total value obtained, in the case when a resource is allocated to the request and in the case when the request is discarded can be expressed as follows:
	The expected value function if $N$ battery units are available and $n$ measurements are targeted, denoted by $V(N,n)$ is expressed as follows:
	\begin{align}
	V(N,n) =
	\left\{
	\begin{array}{ll}
	U(x_i) \mathcal{P}_s^i + \pi \mathbb{E}[V(N,n-1)] +  \\
	\quad (1 - \pi) \mathbb{E}[V(N-1,n-1)], & \mbox{if } x_i \geq a_N^n, \\
	\pi \mathbb{E}[V(N+1,n-1)] +   \\
	\qquad \quad (1-\pi) \mathbb{E}[V(N,n-1)], & \mbox{if } x_i < a_N^n.
	\end{array}
	\right.
	\label{value_func_threshold}
	\end{align}
	\begin{proof}
See \textbf{Appendix~\ref{proof_main_th}}.
	\end{proof}
\end{lemma}


Using this value function, the optimal thresholds for transmission decisions can be obtained using the procedure provided in the following theorem.
\begin{theorem}
If the sensed data valuation at the $i^{\text{th}}$ time slot has valuation $x_i$, the instantaneous channel gain is $h_i$, and the utility achieved by the BS on successful reception of data is $U(x_i)$, then the IoT sensor should transmit the data if its valuation $x_i \geq a_N^n$, where
\begin{align}
a_N^n = U^{-1}&\Bigg[\frac{(1 - \pi)}{\mathcal{P}_s^i} \mathbb{E}[V(N,n-1) - V(N-1,n-1)]  + \notag \\ &\frac{\pi}{\mathcal{P}_s^i} \mathbb{E}[V(N+1,n-1) - V(N,n-1)] \Bigg],
\end{align}
\begin{proof}
See \textbf{Appendix~\ref{proof_main_th_new}}.
\end{proof}
\label{main_th}
\end{theorem}

The expected value function can be obtained using the following lemma.
\begin{lemma}\label{lemma_expect}
	The expected value function if $N$ battery units are available and $n$ measurements are targeted, denoted by $V(N,n)$ is expressed as follows:
	\begin{align}
	\mathbb{E}&[V(N,n)] = \mathbb{E}[\mathcal{P}_s]\int_{a_N^n}^{\infty} U(x)f_X(x) dx + (1 - F_X(a_N^n)) \times \notag \\ &\Bigg[ \pi \mathbb{E}[V(N,n-1)] + (1- \pi)\mathbb{E}[V(N-1,n-1)] \Bigg] + \notag \\
	&F_X(a_N^n) \Bigg[ \pi \mathbb{E}[V(N+1,n-1)] + (1 - \pi)\mathbb{E}[V(N,n-1)] \Bigg].
	\end{align}
	\begin{proof}
		The proof follows from taking the expectation of $V(N,n)$ defined in \textbf{Lemma~\ref{main_th}} with respect to the valuation.
	\end{proof}
\end{lemma}


The evaluation of transmission decision policy requires recursive computations using backward induction. The terminal conditions can be expressed by the following lemma.
\begin{lemma}
	The terminal conditions on the value function expressed in Lemma~\eqref{lemma_value_func} can be expressed as follows:
\end{lemma}
\begin{enumerate}
\item $\mathbb{E}[V(0,n)] = 0, \ \forall n$.\\
  If $N = 0$, there is no battery units left. Hence, the expected value function is zero regardless of the number of target measurements remaining.
  \item $\mathbb{E}[V(N,N)] = N \mathbb{E}[X] \mathbb{E}[\mathcal{P}_s], \ \forall N$.\\
  If there are equal number of battery units as the number of remaining measurements, then the optimal policy is to transmit in all time slots, i.e., $\mathbf{y} = \boldsymbol{1}^N$.
  \end{enumerate}





\subsection{Computation \& Implementation}

In this section, we provide a detailed explanation on the procedure to compute the optimal transmission thresholds by using example scenarios and provide an overview on implementing the proposed framework. The overall process flow is illustrated in Fig.~\ref{fig:flow}. Once the system begins operation, sensing data is collected by the IoT device. The data is then processed to determine how worthy it is to transmit it to the BS. Based on the cutoff thresholds that are precomputed using the statistical information and the developed optimization framework, a decision is made to transmit or to discard the data. The decision is then executed and the system goes back into the earlier state of obtaining the next batch of sensed data until the batteries run out and the operation has ended. To provide an example of the computations involved, we assume a linear utility function on the data valuation, i.e., $U(x) = x$ and obtain analytical expressions for the following special cases:

\begin{algorithm}[H]
	\small
	\caption{Optimal Transmission Policy}
	\begin{algorithmic}[1]
		\Procedure{Threshold Computation}{}
		\For{n = 0 to $n_{\max}$}
		\For{N = 0 to $N_{\max}$}
		\If{$n=0$}
		\State $EV(N,n) \gets 0$
		\ElsIf{$N=0$}
		\State $EV(N,n) \gets \pi EV(N+1,n-1)$
		\ElsIf{$N\ge n$}
		\State $EV(N,n) \gets \mathbb{E}[x]\mathbb{E}[\mathcal{P}_s^i]+\pi EV(N,n-1)+(1-\pi)EV(N-1,n-1)$
		\Else
		\State $T \gets U^{-1}\Bigg[\frac{(1 - \pi)}{\mathbb{E}[\mathcal{P}_s^i]} EV(N,n-1) - EV(N-1,n-1)  +  \frac{\pi}{\mathbb{E}[\mathcal{P}_s^i]} EV(N+1,n-1) - EV(N,n-1) \Bigg]$
		\State $EV(N,n) \gets \mathbb{E}[\mathcal{P}_s]\int_{T}^{\infty} U(x)f_X(x) dx + (1 - F_X(T))\Bigg[ \pi \mathbb{E}[V(N,n-1)] + (1- \pi)\mathbb{E}[V(N-1,n-1)] \Bigg] +F_X(T) \Bigg[ \pi \mathbb{E}[V(N+1,n-1)] + (1 - \pi)\mathbb{E}[V(N,n-1)] \Bigg]$
		\EndIf
		\EndFor
		\EndFor
		\EndProcedure
		\\
		\Procedure{Runtime}{}
		\While{$n\ge0$}
		\State \textit{Collect Data with Value} $x_i$\
		\State \textit{Retrieve Channel Success Probability} $\mathcal{P}_s^i$\
		\State$a_N^n \gets U^{-1}\Bigg[\frac{(1 - \pi)}{\mathcal{P}_s} (EV(N,n-1) - EV(N-1,n-1))  +  \frac{\pi}{\mathcal{P}_s} (EV(N+1,n-1) - EV(N,n-1)] \Bigg]$
		\If{$x_i\ge a_N^n$}
		\State \textit{Transmit Data}
		\State $N \gets N-1$
		\Else
		\State \textit{Discard Data}
		\EndIf
		\State $n \gets n-1$
		\EndWhile
		\EndProcedure
	\end{algorithmic}
	\label{Alg:one}
\end{algorithm}

\subsubsection{Case I: Exponentially distributed data valuation}

If the valuation of sensed data follows an exponential distribution, i.e. $f_X(x) = \Lambda e^{-\Lambda x}$, then the optimal transmission thresholds can be computed according to the following corollary.
\begin{corollary} \label{corollary_exponential}
The decision thresholds if the data valuation is exponentially distributed can be evaluated as follows:
\begin{align}
    a_1^2 &= \frac{(1 - \pi)\mathbb{E}[X]\mathbb{E}[\mathcal{P}_s]}{\mathcal{P}_s^i}= \frac{(1 - \pi)(\alpha_0 + e^{-\mu \rho_{\text{th}}}(\alpha_1 - \alpha_0))}{\Lambda \mathcal{P}_s^i}, \notag \\
    a_1^3 &= \frac{(3\pi-\pi^2)\mathbb{E}[X]\mathbb{E}[\mathcal{P}_s]+(1-2\pi)(1+e^{-2\Lambda a_2^1}(\Lambda a_2^1+1))}{\mathcal{P}_s^i}\notag\\
    &+\frac{e^{-\Lambda a_2^1}(\pi-1)\mathbb{E}[X]\mathbb{E}[\mathcal{P}_s]}{\mathcal{P}_s^i} \notag \\
    &= \frac{(3\pi-\pi^2+e^{-\Lambda a_2^1}(\pi-1))(\alpha_0 + e^{-\mu \rho_{\text{th}}}(\alpha_1 - \alpha_0))}{\Lambda\mathcal{P}_s^i}\notag \\
    &+ \frac{(1-2\pi)(1+e^{-2\Lambda a_2^1}(\Lambda a_2^1+1)}{\mathcal{P}_s^i}
\end{align}
Similarly, the remaining thresholds can be computed iteratively using the terminal conditions.
\begin{proof}
See \textbf{Appendix~\ref{proof_corollarly_exponential}}.
\end{proof}
\end{corollary}

\subsubsection{Case II: Uniformly distributed data valuation}
If the valuation of sensed data follows a uniform distribution, i.e. $f_X(x) = \frac{1}{\overline{x}-\underline{x}}$, then the optimal transmission thresholds can be computed according to the following corollary.

\begin{corollary}
The decision thresholds if the data valuation is uniformly distributed can be evaluated as follows:
\begin{align}
    a_1^2 &=  \frac{(1 - \pi)\mathbb{E}[X]\mathbb{E}[\mathcal{P}_s]}{\mathcal{P}_s^i}, \notag \\
    &= \frac{(1 - \pi)(\bar{x}+\underline{x})(\alpha_0 + e^{-\mu \rho_{\text{th}}}(\alpha_1 - \alpha_0))}{2 \mathcal{P}_s^i}, \notag \\
    a_1^3 &= \frac{(-4\pi^2+5\pi+1)(\bar{x}+\underline{x})(\alpha_0 + e^{-\mu \rho_{\text{th}}}(\alpha_1 - \alpha_0))}{4\mathcal{P}_s^i} \notag \\
    &+ \frac{(3\bar{x}^2+2\bar{x}\underline{x}+\underline{x}^2)(1-2\pi)(\alpha_0 + e^{-\mu \rho_{\text{th}}}(\alpha_1 - \alpha_0))}{16(\bar{x}-\underline{x})\mathcal{P}_s^i}
\end{align}
\end{corollary}

The remaining thresholds for arbitrary number of remaining measurements and battery units can be recursively computed. The step-wise procedure for computing the thresholds and the real-time implementation procedure is summarized in Algorithm~\ref{Alg:one}.



\vspace{-0.0in}
\section{Results} \label{Sec:Results}
In this section, we provide results based on the numerical evaluation of the performance of our proposed dynamic decision framework. For the sake of illustration, we consider two special cases, i.e., exponentially and uniformly distributed data valuation with and without energy harvesting. For illustrative purposes, we select the model parameters as follows: the duration of each time slot is selected to be $\tau = 10$ ms, the average probability of successful transmission in good and bad channel states are set as $\alpha_1 = 0.8$ and $\alpha_0 = 0.2$ respectively, the threshold on channel gain is set to be $\rho_{th} = 0.5$, and the channel gain parameter is selected as $\mu = 0.5$.

In order to test the efficiency of our mechanism, we run several numerical tests and compare it with existing protocols.
The metrics we used for evaluating the performance are total utility gained and the achieved battery lifetime. For each test, we random draw $1000$ data valuation samples, referred to as $x_1, x_2, \dots, x_{1000}$ according to their distribution.
We assume that the sensor has $N$ battery units available, where $N = 1, 2, \dots, 100$. Each sensor follows a fixed strategy. At the $i$-th stage, each sensor decides on whether to transmit the data based on its strategy until it runs out of battery or the test ends. Then we compare the total utility gained and battery lifetime. For each data valuation distribution and energy harvesting condition, the tests are repeated $100$ times to average out the effect of randomness.

\begin{figure*}[t]
	\centering
	\subfloat[Data valuations sampled from an exponential distribution.]{\includegraphics[width=3.5in]{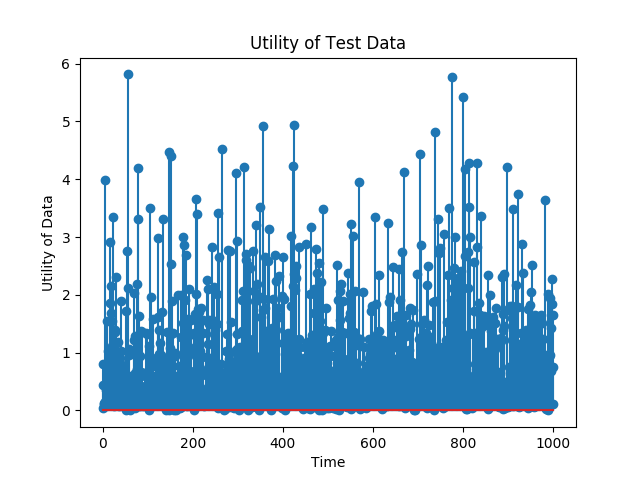} \label{fig:trace_exp_0}}
	\subfloat[Data valuations sampled from a uniform distribution.]{\includegraphics[width=3.5in]{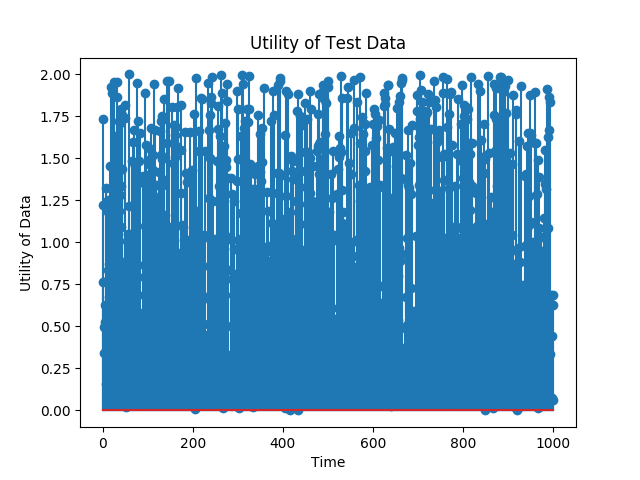} \label{fig:trace_uni_0}}
	\caption{Example traces of data valuations used in experimental results.}
	\label{results}
\end{figure*}


\begin{figure*}[t]
	\centering
	\subfloat[Optimal Threshold for transmitting data with exponentially distributed data valuation and no energy harvesting, i.e.,with $\pi=0$.]{\includegraphics[width=3.5in]{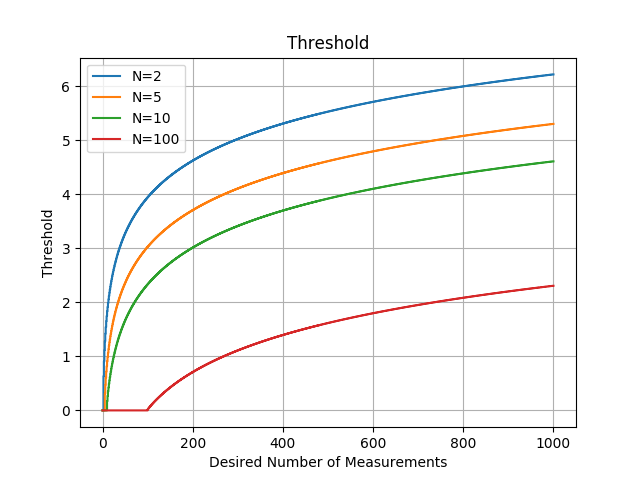} \label{fig:thr_exp_0}}
	\subfloat[Optimal Threshold for transmitting data with uniformly distributed data valuation and no energy harvesting, i.e., $\pi=0$.]{\includegraphics[width=3.5in]{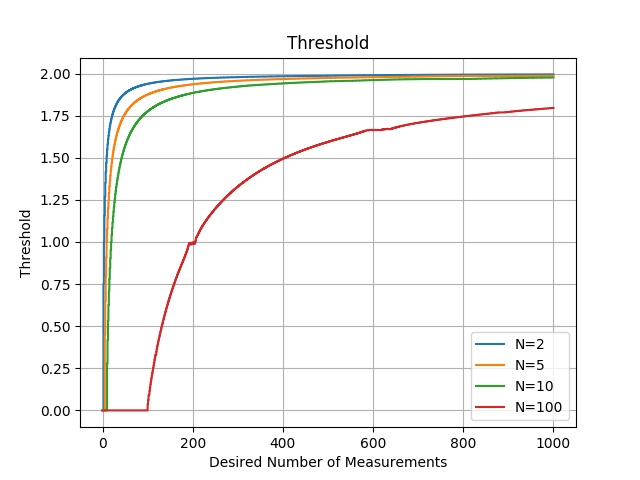}  \label{fig:thr_uni_0} }
	\caption{Traces of data valuation used in experimental results.}
	\label{results}
\end{figure*}

\subsection{Exponential Distribution with No Energy Harvesting}
In this case, the distribution of data valuation and the energy harvesting condition is assumed to follow an exponential distribution with unit mean, i.e. $\lambda=1$. The pdf of data valuation can be expressed as follows:
\begin{equation}
    f_X(x) =
    \begin{cases}
        e^{-x} &\text{if } x\ge0\\
        0 &\text{otherwise}
    \end{cases}
    \text{, }\pi = 0
\end{equation}
A trace of the randomly drawn data valuations is shown in Figure ~\ref{fig:trace_exp_0}. The optimal transmission thresholds for varying battery levels and target measurements are plotted in Fig.~\ref{fig:thr_exp_0}. It can be observed that the thresholds increase as the target number of measurements are higher. This is because as more target measurements are remaining, there is a possibility of higher valued data to arrive. However, as fewer target measurements are remaining, the thresholds are lower, i.e., the device is less selective in transmitting data to achieve more utility.

\subsection{Uniform Distribution with No Energy Harvesting}
In this case, the distribution of data valuation and the energy harvesting condition is:
\begin{equation}
f_X(x) =
\begin{cases}
0.5 &\text{if } 0\le x\le 2\\
0 &\text{otherwise}
\end{cases}
\text{, }\pi = 0
\end{equation}

A trace of the randomly drawn data valuations is shown in Figure \ref{fig:trace_uni_0}.
The optimal thresholds in this case are illustrated in Fig.~\ref{fig:thr_uni_0}. We plotted the optimal thresholds for varying battery levels and target measurements. It can observed that as the expected number of measurements increases, the threshold gets higher, which means that the optimal algorithm holds the battery and waits for higher valued data when there are more target measurements. When the sensor has more battery units available, it lowers the threshold to achieve maximum total utility. For smaller $N$, the thresholds converge to $x_{\max}=2$ very quickly because transmitting data is very costly the chance of a data with valuation close to $2$ is very high. For any battery level, when the number of measurements is less than the number of battery units, which is $n\le N$, there is sufficient battery and the optimal transmission policy is to transmit the sensed data, and thus the optimal threshold is zero.



\subsection{Exponential Distribution with Energy Harvesting}
In this test we examine the effect of energy harvesting. The updated inputs with the probability of harvesting energy, $\pi = 0.1$, are expressed as follows:
\begin{equation}
    f_X(x) =
    \begin{cases}
        e^{-x} &\text{if } x\ge0\\
        0 &\text{otherwise}
    \end{cases}
    \text{, }\pi = 0.1
\end{equation}

\begin{figure}
	\centering
	\includegraphics[width=3.5in]{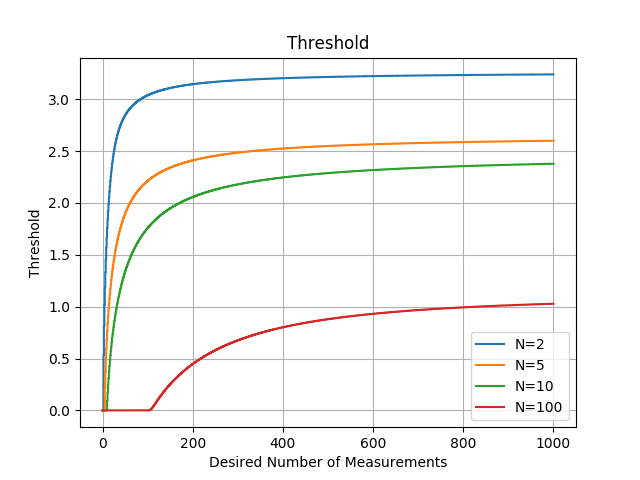}
	\caption{Optimal threshold for exponentially distributed data valuations with $\pi=0.1$.}
	\label{thr_exp_harvest}
\end{figure}

\begin{figure}[h]
	\includegraphics[width=3.5in]{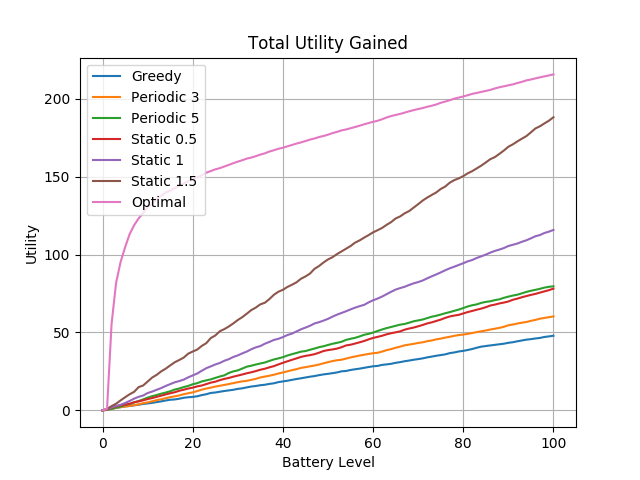}
	\caption{Average Total Utility for Exponentially Distributed Data with $\pi=0.1$.}
	\label{fig:uti_exp_0_1}
\end{figure}

The optimal transmission thresholds are shown in Fig.~\ref{thr_exp_harvest}. In comparison to the test with no energy harvesting, the optimal thresholds under energy harvesting are lower. Intuitively, energy harvesting allows more battery units to be available for transmitting, and therefore the curves have the same trend but are lower implying less selectivity. Fig.~\ref{fig:uti_exp_0_1} shows the average total utility. It is interesting to note that the optimal policy gains much more utility than the others when fewer battery units are available, and the gap decreases as battery level increases. This can be explained by the pattern in the thresholds. With small $N$, the threshold increases very fast as battery increases. But the threshold curves saturate with very large $N$ and the utility curves thus become straight lines.


\begin{figure}
	\centering
	\includegraphics[width=3.5in]{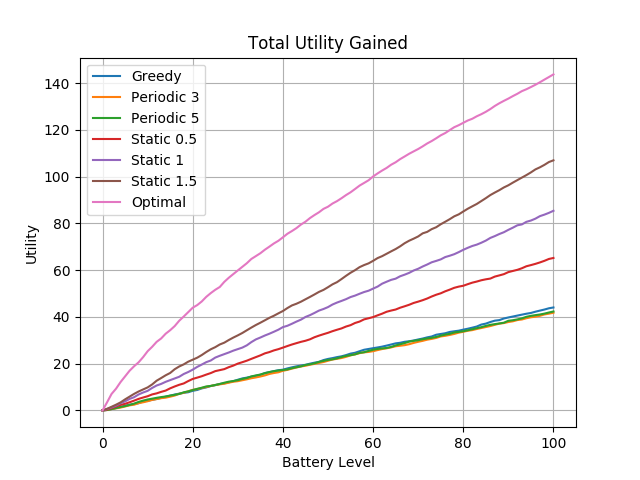}
	\caption{Average total utility for exponentially distributed\\ data valuations with $\pi=0$.}
	\label{fig:uti_exp_0}
\end{figure}

\begin{figure}
	\centering
	\includegraphics[width=3.5in]{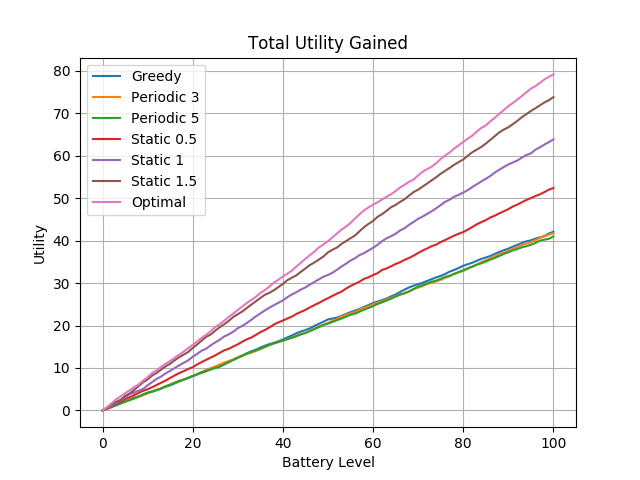}
	\caption{Average total utility for uniformly distributed data with $\pi=0$.}
	\label{fig:uti_uni_0}
\end{figure}


\begin{figure*}[h]
	\centering
	\subfloat[Average Battery Lifetime for Exponentially Distributed Data with $\pi=0$.]{\includegraphics[width=3.5in]{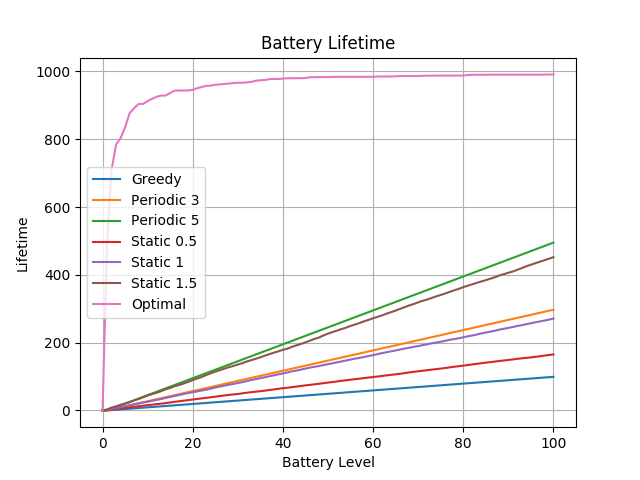} \label{fig:bat_exp_0}}
	\subfloat[Average Battery Lifetime for Uniformly Distributed Data with $\pi=0$.]{\includegraphics[width=3.5in]{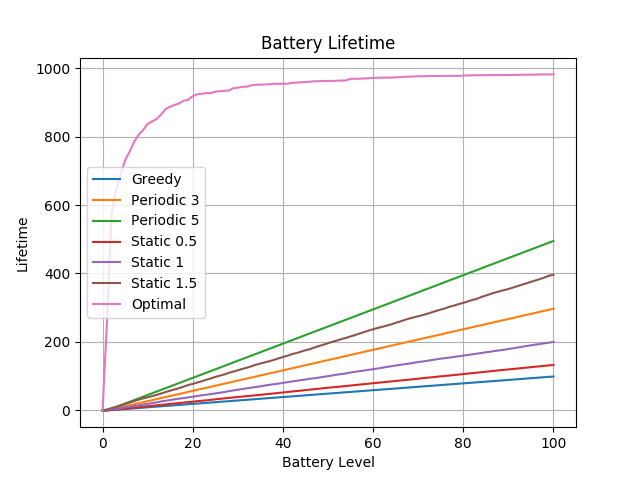} \label{fig:bat_uni_0}}
	\caption{Traces of data valuation used in experimental results.}
	\label{results}
\end{figure*}


\subsection{Comparison and Numerical Experiments}

In this section, we compare our proposed optimal decision framework with other benchmark strategies.
The main strategies used for comparison are as follows:
\begin{enumerate}
    \item \textbf{Greedy Transmission:}
    In this strategy, the sensors transmit the data as soon as they sense it when battery is available regardless of the valuation of sensed data. In other words, this is a non-strategic transmission strategy.
    \item \textbf{Periodic Transmission:} This policy is a sleep-wake cycle inspired policy that does not use any intelligence in transmitting data. Instead, it transmits data periodically regardless of the valuation of data. In our experiments, we use a period of 3 and 5 time slots which implies that a regular transmission is made in every third and fifth time slot respectively.
    \item \textbf{Static Threshold:} In this transmission strategy, a transmission is made only once the valuation has crossed a set barrier. In other words, the data is transmitted if its valuation determined by the sensor is higher than a particular level. Several different levels are investigated as part of our comparison.
\end{enumerate}

Fig.~\ref{fig:uti_exp_0} and Fig.~\ref{fig:uti_uni_0} plots the average total utility against different starting battery levels for exponential and uniformly distributed data valuations. It is obvious that the curve for optimal policy is not linear any more because the threshold does not converge, while the rest of the strategies remain linear. And therefore the optimal strategy receives more utility by a higher margin. Fig.~\ref{fig:uti_exp_0_1} shows the average utility under energy harvesting. It can be observed that the presence of energy harvesting significantly enhances the average utility obtained by the BS.
In comparison with other strategies, Fig.~\ref{fig:bat_exp_0} and Fig.~\ref{fig:bat_uni_0} shows the average battery time for different strategies with different starting battery level in the case of exponential and uniform data valuations respectively. The battery lifetime shown in Fig.~\ref{fig:bat_uni_0} shows that the optimal policy achieves a huge gain in operation time when the battery is insufficient. The optimal policy reaches almost $1000$ with very limited starting battery level, which means that it can almost work until the desired operation time ends with few battery while gains maximal utility. Fig.~\ref{fig:bat_exp_0} illustrates the average battery lifetime in the exponential case is almost identical to the uniform case where the optimal policy wins with a large gap ahead of the comparison policies.

\begin{figure}
    \centering
    \includegraphics[width=3in]{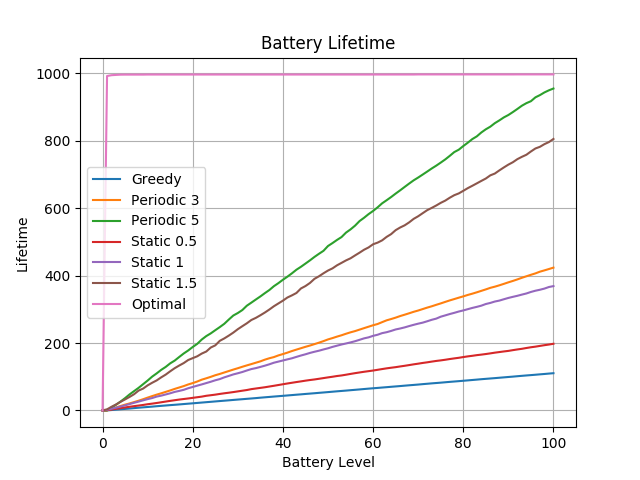}
    \caption{Average Battery Lifetime for Exponential Distributed Data with $\pi=0.1$}
    \label{fig:bat_exp_0.1}
\end{figure}

The battery lifetime under energy harvesting scenario is displayed in Figure \ref{fig:bat_exp_0.1}. With energy harvesting, the gap between the optimal policy and the comparison strategies is huge from the beginning. When battery is very limited, the optimal policy waits longer than the ones with static thresholds and thus receive the full benefits of energy harvesting. Policies with low threshold are shut down very early and no further harvested energy is gained.

The results demonstrate that there can be significant gains in the performance as well as operation lifetime of the IoT sensors using the developed transmission framework. However, these numbers depend on the specific utility function employed as well as the target number of measurements set by the sensor. Hence, the gains cannot be generalized for arbitrary cases.

\vspace{-0.0in}
\section{Conclusion \& Future Work} \label{Sec:Conclusion}
In this paper, we develop an online transmission mechanism under power saving mode for IoT sensor devices. A dynamic programming approach has been used to optimally make real-time decisions regarding transmission of sensed data based on instantaneous channel conditions and battery units available. The availability of energy harvesting and its impact on transmission policies has also been investigated.
This work is a starting point in developing an intelligent transmission protocol that uses the valuation of the data in transmission decisions. We expect that such cross-layer protocols will pave the way for the development of more cognitive and strategic transmission mechanisms and to enhance the operational lifetime of battery powered devices while achieving mission specific goals. In practice, if the statistics of data valuation can be accurately predicted, there is a potential for significant improvement in the utility as well as battery lifetime of the sensor.

Several different directions can be pursued as part of the future work. The case where the data valuation at each time step depends on the valuation of the previous one can be explored. A markov chain modeling is required to model the data valuation representing different application scenarios.
\appendices

\vspace{-0.0in}
\section{Proof of Lemma~\ref{main_th}} \label{proof_main_th}
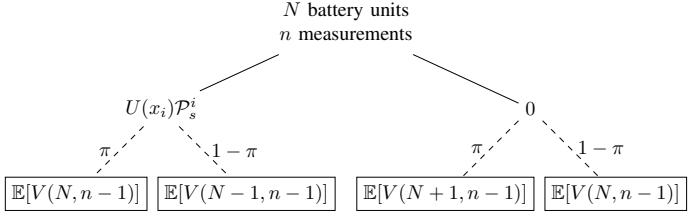
\begin{figure}
\centering
\begin{adjustbox}{width=0.5\textwidth}
\begin{tikzpicture}[level distance=1.5cm,
  level 1/.style={sibling distance=6.5cm},
  level 2/.style={sibling distance=3cm},scale = 1]
  \node {\begin{tabular}{c} $N$ battery units \\ $n$ measurements \end{tabular}}
    child {node {$U(x_i) \mathcal{P}_s^i$}
      child[dashed] {node[draw,solid] {$\mathbb{E}[V(N,n-1)]$} edge from parent node[left] {$\pi$} }
      child[dashed] {node[draw,solid] {$\mathbb{E}[V(N-1,n-1)]$} edge from parent node[right] {$1 - \pi$}}
    }
    child {node {$0$}
      child[dashed] {node[draw,solid] {$\mathbb{E}[V(N+1,n-1)]$} edge from parent node[left] {$\pi$} }
      child[dashed] {node[draw,solid] {$\mathbb{E}[V(N,n-1)]$} edge from parent node[right] {$1 - \pi$}}
    };
\end{tikzpicture}
\end{adjustbox}
\caption{Decision tree for the $i^{\text{th}}$ time slot if $N$ battery units are available and $n$ measurements are remaining.}
\label{fig:decision_tree}
\end{figure}

For the general case, if the sensor has $N \geq 1$ units of battery remaining and $n > n$ target measurements at the $i^{\text{th}}$ time slot, then the decision has to be made to transmit the current measurement $x_i$ or skip transmission in the hope of being able to transmit data with higher valuation in the future. If the data is transmitted, a utility $U(x_i)$ will be received at the BS with a probability $\mathcal{P}_s^i$, i.e., if the data successfully gets transmitted, and $N-1$ battery units are remaining for future $n-1$ measurements. On the other hand, if the decision is to skip the transmission, then there is zero utility at the BS with certainty and $N$ battery units are still available for the remaining $n-1$ measurements. However, once the decision has been taken at the $i^{\text{th}}$ time slot, there might be an additional battery unit available for transmission from the harvested energy with a probability of $\pi$. Therefore, there is an addition of a battery unit with a probability of $\pi$ making $N$ and $N+1$ battery units available in the cases when the sensor transmits and skips respectively. The decision problem is illustrated in the form of a tree in Figure~\ref{fig:decision_tree} where the leaves point towards the expected future utility with the remaining battery units and target measurements after the initial decision has been made. Consequently, at the root node, the value function if $N$ battery units and $n$ transmissions are remaining can be expressed as follows:
 \begin{align}
    &V(N,n)  = \underset{y \in \{0,1\}}{\max} \Big\lbrace
             y \left( U(x_{i})\mathcal{P}_s^i + \pi \mathbb{E}[V(N,n-1)] + \right. \notag \\& \left. (1-\pi)\mathbb{E}[V(N-1,n-1)] \right), (1-y)\left(
             \pi \mathbb{E}[V(N+1,n-1)] +  \right. \notag \\ & \left. (1-\pi) \mathbb{E}[V(N,n-1)] \right) \Big\rbrace,
             \label{value_func_N_n}
\end{align}
where the variable $y = 1$ implies that the sensor should transmit at the current time slot, while $y = 0$ implies that the transmission should be skipped. The solution to~\eqref{value_func_N_n} is $y = 1$, i.e., to transmit, if $U(x_{i})\mathcal{P}_s^i + \pi \mathbb{E}[V(N,n-1)] + (1-\pi)\mathbb{E}[V(N-1,n-1)] \geq \pi \mathbb{E}[V(N+1,n-1)] +  (1-\pi) \mathbb{E}[V(N,n-1)]$ and $y = 0$ otherwise, i.e., not transmit. It leads to the transmission condition in Theorem~\ref{main_th} based on the instantaneous data valuation $x_i$ and transmission success probability $\mathcal{P}_s^i$. We denote the comparison threshold on $x_i$ when $N$ battery units are available and $n$ measurements are remaining as $a_N^n$. It is then straightforward to show that the value function in~\eqref{value_func_N_n} can be expressed as in~\eqref{value_func_threshold}.

\section{Proof of Theorem~\ref{main_th}} \label{proof_main_th_new}
From the decision tree shown in Fig.~\ref{fig:decision_tree}, the a transmission is only made if the average utility obtained is higher than postponing it for future measurements. It implies that
\begin{align}
U(x_i) \mathcal{P}_s^i + \pi \mathbb{E}[V(N,n-1)] +  (1 - \pi) \mathbb{E}[V(N-1,n-1)] \geq   \notag \\
\pi \mathbb{E}[V(N+1,n-1)] +  (1-\pi) \mathbb{E}[V(N,n-1)].
\end{align}
This leads to the condition that
\begin{align}
x_i    \geq  \frac{1}{\mathcal{P}_s^i}  U^{-1}  \Bigg[ \pi \left( \mathbb{E}[V(N+1,n-1)] -  \mathbb{E}[V(N,n-1)] \right) + \notag  \\   (1-\pi) \left( \mathbb{E}[V(N,n-1)] - \mathbb{E}[V(N-1,n-1)]   \right)  \Bigg].
\end{align}
The expectation of the value function can be obtained directly from the definition in Lemma~\eqref{lemma_value_func}.

\section{Proof of Corollary~\ref{corollary_exponential}}\label{proof_corollarly_exponential}
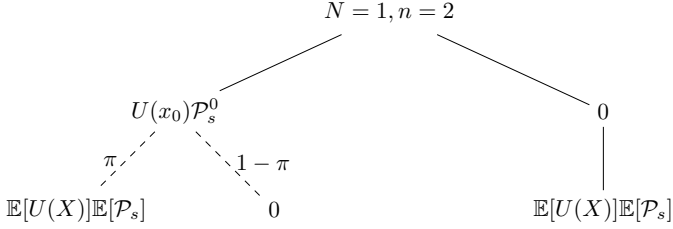
\begin{figure}
    \centering
 \begin{adjustbox}{width=0.5\textwidth}
\begin{tikzpicture}[level distance=1.5cm,
  level 1/.style={sibling distance=6.5cm},
  level 2/.style={sibling distance=3cm},scale = 1]
  \node {\begin{tabular}{c} $N=1, n = 2$ \end{tabular}}
    child {node {$U(x_{0}) \mathcal{P}_s^{0}$}
      child[dashed] {node[solid] {$\mathbb{E}[U(X)]\mathbb{E}[\mathcal{P}_s]$} edge from parent node[left] {$\pi$} }
      child[dashed] {node[solid] {$0$} edge from parent node[right] {$1 - \pi$}}
    }
    child {node {$0$}
      child {node[solid] {$\mathbb{E}[U(X)]\mathbb{E}[\mathcal{P}_s]$} }
    };
\end{tikzpicture}
\end{adjustbox}
    \caption{Decision tree at $i^{\text{th}}$ time slot with $N=1$ and $n=2$.}
    \label{fig:decision_tree_N1_n2}
\end{figure}

If there are two measurements remaining to achieve the target and one unit of battery, i.e., $N=1$ and $n = 2$, then at the measurement of data at the $(n-2)^{\text{th}}$ time slot, the decision tree is shown in Figure~\ref{fig:decision_tree_N1_n2}.
The value function is expressed as follows:
\begin{align}
V(2,1) =
\left\{
	\begin{array}{ll}
		U(x_i) \mathcal{P}_s^i + \pi \mathbb{E}[U(X)] \mathbb{E}[\mathcal{P}_s],  & \mbox{if } x_i \geq a_2^1, \\
		\mathbb{E}[U(X)] \mathbb{E}[\mathcal{P}_s], & \mbox{if } x_i < a_2^1.
	\end{array}
\right.
\end{align}
The sensor should transmit data if
\begin{align}
x_{i} > a_2^1 =  U^{-1}\Bigg[\frac{(1 - \pi)}{\mathcal{P}_s^i} \mathbb{E}[U(X)] \mathbb{E}[\mathcal{P}_s] \Bigg],
\end{align}

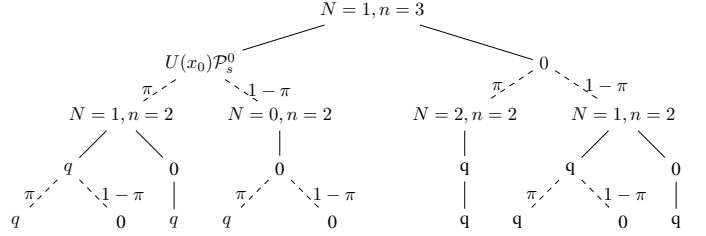
\begin{figure}
    \centering
 \begin{adjustbox}{width=0.5\textwidth}
\begin{tikzpicture}[level distance=1cm,
  level 1/.style={sibling distance=6.5cm},
  level 2/.style={sibling distance=3cm},
  level 3/.style={sibling distance=2cm}]
  \node {$N=1, n = 3$ }
    child {node {$U(x_{0}) \mathcal{P}_s^{0}$}
      child[dashed] {node[solid] {$N=1, n=2$} child[solid] {node {$q$} child[dashed] {node {$q$} edge from parent node[left] {$\pi$}} child[dashed] {node {0} edge from parent node[right] {$1 - \pi$}} } child[solid] {node {0} child[solid] {node {$q$}}} edge from parent node[left] {$\pi$} }
      child[dashed] {node[solid] {$N=0, n=2$} child[solid] {node {0} child[dashed] {node {$q$} edge from parent node[left] {$\pi$}} child[dashed] {node {0} edge from parent node[right] {$1 - \pi$}}} edge from parent node[right] {$1 - \pi$}}
    }
    child {node {$0$}
        child[dashed] {node {$N=2, n=2$} child[solid] {node {q} child[solid] {node {q} } } edge from parent node[left] {$\pi$}}
        child[dashed] {node {$N=1, n=2$} child[solid] {node {q} child[dashed] {node {q} edge from parent node[left] {$\pi$}} child[dashed] {node {0} edge from parent node[right] {$1 - \pi$} } } child[solid] {node {0} child[solid] {node {q}} } edge from parent node[right] {$1 - \pi$}}
    };
\end{tikzpicture}
\end{adjustbox}
    \caption{Decision tree at $i^{\text{th}}$ time slot with $N=1$ and $n=3$. For notational convenience, $q = \mathbb{E}[U(X)]\mathbb{E}[\mathcal{P}_s]$.}
    \label{fig:decision_tree_N1_n3}
\end{figure}
If there are three measurements remaining, then the decision tree is three levels deep as shown in Figure~\ref{fig:decision_tree_N1_n3}. The value function is expressed as follows:
\begin{align}
V(3,1) =
\left\{
	\begin{array}{ll}
		U(x_i) \mathcal{P}_s^i + \pi \mathbb{E}[V(2,1)]  + (1-\pi) \mathbb{E}[V(2,0)] , \\
 &      \hspace{-1.5in}  \mbox{if } x_i \geq a_3^1, \\
		\pi \mathbb{E}[V(2,2)]  + (1-\pi) \mathbb{E}[V(2,1)] , \\
 &      \hspace{-1.5in} \mbox{if } x_i < a_3^1.
	\end{array}
\right.
\end{align}

It follows that a transmission should be made if:
\begin{align}
    x_{i} > a_3^1 = U^{-1}\Bigg[\frac{\pi}{\mathcal{P}_s^i} (\mathbb{E}[V(2,2)]-\mathbb{E}[V(1,2)]) \notag\\+ \frac{1-\pi}{\mathcal{P}_s^i}(\mathbb{E}[V(1,2)]-\mathbb{E}[V(0,2)])\Bigg]\notag.
\end{align}
By computing the expected value functions using Lemma~\ref{lemma_expect}, we get the desired threshold.


\vspace{-0.0in}
\bibliographystyle{IEEEtran}
\bibliography{references2}

\end{document}